\documentclass[11pt,a4paper]{article}

\usepackage[a4paper,margin=2.2cm]{geometry}
\usepackage{amsmath,amssymb,graphicx,cite}
\usepackage{hyperref}

\begin{document}

\title{Spherical thin shells with charge in unimodular gravity}

\author{Gabriel R. Bengochea\thanks{e-mail: gabriel@iafe.uba.ar}, Ernesto F. Eiroa\thanks{e-mail: eiroa@iafe.uba.ar}, Griselda Figueroa-Aguirre\thanks{e-mail: gfigueroa@iafe.uba.ar}\\
{\small  Instituto de Astronom\'{\i}a y F\'{\i}sica del Espacio (IAFE, CONICET-UBA),}\\
{\small Ciudad Universitaria, 1428, Buenos Aires, Argentina}} 

\date{}

\maketitle

\begin{abstract}
We construct spherically symmetric thin shells in unimodular gravity (UG) and analyze their dynamical stability under radial, symmetry-preserving perturbations. We use the UG junction conditions and adopt non-conservative charged exterior solutions sourced by a radial electric field. As an application, we build charged vacuum bubbles: a Minkowski interior surrounded by a charged thin shell and matched to an exterior UG geometry. The resulting spacetimes are free of event horizons and central singularities, while in de Sitter-like cases a cosmological horizon may exist outside the shell. We compare the stability regions and the matter content with their counterparts in general relativity, finding some differences controlled by the UG non-conservation parameter. This work extends previously published results.
\end{abstract}

\section{Introduction}

Unimodular gravity (UG) is a formulation of the gravitational dynamics closely related to general relativity (GR), but based on a different treatment of the spacetime volume element. In this framework, the metric volume form is constrained by a fixed non-dynamical four-volume element, so that the gravitational field equations can be written in trace-free form. One important consequence is that the cosmological constant is not introduced as a prescribed coupling in the field equations, but appears as an integration constant fixed by initial conditions. This feature has motivated considerable interest in UG in connection with the cosmological constant problem, since contributions to the energy--momentum tensor proportional to the metric, such as vacuum-energy terms, do not enter the trace-free equations in the same way as in GR \cite{weinberg,carroll,ellis}.

Another relevant aspect of UG is that the usual covariant conservation of the matter energy--momentum tensor is not forced by the trace-free gravitational equations alone. If standard conservation is imposed as an additional assumption, UG reproduces the local content of GR with a cosmological constant. However, if this assumption is relaxed, the non-conservation can be parametrized through a diffusion-like scalar contribution, and the effective cosmological term becomes spacetime dependent \cite{ellis}. In that situation, UG may lead to genuine departures from GR. This possibility has been explored in different contexts, including cosmology, see for example \cite{ellis14,corral20,josset,fabris19,cesare22,leon22,bengo23}, compact objects \cite{astorga19}, and static spherically symmetric geometries. In this work, we are interested in two families of static charged UG geometries inspired by the spherically symmetric solutions with a radial electric field obtained in \cite{fabris}. As shown in \cite{fabris,BEFA2025}, in the absence of the non-conservative sector, the exterior geometry reduces to the Reissner--Nordstr\"om solution of GR with a cosmological constant.

Thin shells are a useful tool for constructing idealized compact configurations in which matter is localized on a hypersurface separating two spacetime regions. In GR, the Darmois--Israel formalism relates the jump of the extrinsic curvature across the hypersurface to the surface energy--momentum tensor, allowing one to determine the shell energy density and the pressure, and also its dynamics \cite{israel1966}. This technique has been widely used to model vacuum bubbles, shells around compact objects, wormholes, and gravastar-like configurations, e.g. \cite{brady91,poisson95,ishak02,eiroa04,gravstar04,lobo05,eiroa08,eiroa11,montelongo12,forgani18,eiroa19,berry20}. For UG, the junction conditions introduced in \cite{ellis} have a trace-free structure inherited from the corresponding field equations. In \cite{BEFA2025}, these conditions were applied to spherical thin shells built from non-conservative charged exterior geometries, making it possible to analyze how the UG non-conservative sector affects the shell matter content and its radial stability.

In this article, we summarize the construction of spherically symmetric charged thin shells in UG, following the analysis of \cite{BEFA2025}. We consider a Minkowski interior joined to a non-conservative exterior solution with a radial electric field, obtaining charged vacuum bubbles without event horizons or central singularities. We then discuss the linear stability of the static configurations under radial perturbations and compare the resulting stability regions with the GR counterpart, namely the Reissner--Nordstr\"om geometry with a cosmological constant. The aim is to highlight, in a compact way, how the UG non-conservation parameter affects the critical charge, the weak energy condition at the shell, and the domain of stable configurations. We present some new results not found in the literature. We adopt units such that $G=c=1$.

\section{UG and non-conservative charged solutions}

A convenient expression for the UG action is \cite{weinberg,carroll}
\begin{equation}
S[g_{ab},\Psi_M;\Lambda]=\frac{1}{2\kappa}\int \left( R\,\epsilon^{(g)}_{abcd}-2\Lambda\left(\epsilon^{(g)}_{abcd}-\varepsilon_{abcd}\right)\right) +S_M[g_{ab},\Psi_M],
\label{eq:action}
\end{equation}
with $\kappa\equiv 8\pi$, $\varepsilon_{abcd}$ a fixed 4-volume element, and $\epsilon^{(g)}_{abcd}$ the metric 4-volume element.
The field equations can be written in the trace-free form
\begin{equation}
R_{ab}-\frac14 g_{ab}R=\kappa\left(T_{ab}-\frac14 g_{ab}T\right),
\label{eq:ug_tracefree}
\end{equation}
where $R_{ab}$ is the Ricci tensor and $T_{ab}$ is the energy--momentum tensor, while $R$ and $T$ are their respective traces. UG allows a generalized non-conservation law
\begin{equation}
\nabla_a\left(T^{ab}-g^{ab}D\right)=0,
\label{eq:noncons}
\end{equation}
where $D(x)$ is an arbitrary scalar function. As a result,
\begin{equation}
\Lambda(x)=\Lambda_0+\kappa D(x),
\label{eq:Lambda}
\end{equation}
with $\Lambda_0$ an integration constant fixed by initial conditions.

The line element for the non-conservative static and spherically symmetric solutions, in the presence of a radial electric field, reads 
\begin{equation}
ds^2=-A(r)\,dt^2+A(r)^{-1}dr^2+r^2(d\theta^2+\sin^2\theta\,d\phi^2),
\label{eq:sss_metric}
\end{equation}
where $A(r)$ for two representative families has been recently found \cite{fabris,BEFA2025}.
\paragraph{Case A:}
The choice 
\begin{equation}
\Lambda(r)=\Lambda_0+\Lambda_1 r^{p}, \label{eq:caseA_Lambda}
\end{equation}
for $p\neq \{-4,-3\}$, leads to
\begin{align}
A(r)&=1-\frac{2M}{r}+\frac{Q^2}{r^2}-\frac{\Lambda_0}{3}r^2-\frac{4\Lambda_1}{(p+3)(p+4)}r^{p+2}, \label{eq:caseA_A}\\
E^2(r)&=\frac{Q^2}{r^4}-\frac{p}{p+4}\Lambda_1 r^{p}, \label{eq:caseA_E}
\end{align}
with $M$ the mass, $Q$ the electric charge, and $\Lambda _1 $ a constant. The GR Reissner--Nordstr\"om--(A)dS solution is recovered for $p=0$ after $\Lambda_0\to \Lambda_0+\Lambda_1$. When $p=-2$, after an appropriate change of coordinates, the metric function can be rearranged so that the spacetime presents an angular deficit or surplus \cite{BEFA2025}. We are not interested in the solutions for $p = \{-4,-3\}$, which can be found in \cite{fabris,BEFA2025}.
\paragraph{Case B:}
Another choice
\begin{equation}
\Lambda(r)=\Lambda_0+\frac{\Lambda_1}{(r^2+b^2)^2}, \label{eq:caseB_Lambda}
\end{equation}
results in
\begin{align}
A(r)&=1-\frac{2M}{r}+\frac{Q^2}{r^2}-\frac{\Lambda_0}{3}r^2  
-\frac{2\Lambda_1}{r^2}\left(\frac{r}{b}\arctan\!\left(\frac{r}{b}\right)-\ln\!\left(1+\frac{r^2}{b^2}\right)\right), \label{eq:caseB_A}\\
E^2(r)&=\frac{Q^2}{r^4}+\frac{\Lambda_1}{r^4}\left(\frac{b^2(3b^2+4r^2)}{(b^2+r^2)^2}+2\ln\!\left(1+\frac{r^2}{b^2}\right)\right), \label{eq:caseB_E}
\end{align}
where $b$ is a constant. In both scenarios, the spacetime represents a black hole with a singularity at the center as long as the charge does not exceed the extremal value $Q_c$, above which the event horizon vanishes and the singularity becomes naked \cite{fabris,BEFA2025}. A cosmological horizon is also present when $\Lambda_0 >0$. For $\Lambda_1=0$ both cases reduce to the GR charged solution with cosmological constant $\Lambda_0$.

\section{Spherical thin shells in UG: junction conditions and stability}

We join two manifolds with metrics \eqref{eq:sss_metric}, characterized by functions $A_1(r)$ (inner) and $A_2(r)$ (outer), across a timelike hypersurface $\Sigma$ defined by $r=a(\tau)$ in order to construct a new one with a thin shell at  $\Sigma$.
The UG junction conditions read \cite{ellis,BEFA2025}
\begin{equation}
-\left[K_{\mu\nu}-K\left(h_{\mu\nu}-\frac12 g_{\mu\nu}\right)\right]
=8\pi\left(S_{\mu\nu}-\frac14 S g_{\mu\nu}\right),
\label{eq:junction}
\end{equation}
where $h_{\mu\nu}$ is the induced metric on $\Sigma$, $K_{\mu\nu}$ the extrinsic curvature, $K$ its trace, and $S_{\mu\nu}$ the surface energy--momentum tensor, while the brackets denote the jump across $\Sigma$.

Assuming a perfect fluid on the shell, $S_{\hat\imath\hat\jmath}=\mathrm{diag}(\sigma,p,p)$ in an orthonormal basis, the surface density and transverse pressure are
\begin{align}
\sigma &= -\frac{1}{4\pi a}\left(\sqrt{A_2(a)+\dot a^2}-\sqrt{A_1(a)+\dot a^2}\right), \label{eq:sigma_dyn}\\
p&= -\frac{\sigma}{2}+\frac{1}{16\pi}\left(\frac{2\ddot a + A_2'(a)}{\sqrt{A_2(a)+\dot a^2}}-\frac{2\ddot a + A_1'(a)}{\sqrt{A_1(a)+\dot a^2}}\right), \label{eq:p_dyn}
\end{align}
where the prime and the dot denote the derivatives with respect to the radial coordinate and the proper time $\tau$ at the shell, respectively. These equations, or any of them and
\begin{equation}
\frac{d(a^2\sigma)}{d\tau}+p\frac{da^2}{d\tau}=0,
\label{conserv}
\end{equation}
determine $a(\tau)$. For static configurations with shell radius $a=a_0$ we obtain 
\begin{align}
\sigma_0 &= -\frac{1}{4\pi a_0}\left(\sqrt{A_2(a_0)}-\sqrt{A_1(a_0)}\right), \label{eq:sigma0}\\
p_0&= -\frac{\sigma_0}{2}+\frac{1}{16\pi}\left(\frac{A_2'(a_0)}{\sqrt{A_2(a_0)}}-\frac{A_1'(a_0)}{\sqrt{A_1(a_0)}}\right). \label{eq:p0}
\end{align}
The shell dynamics can be written as
\begin{equation}
\dot a^2+V(a)=0,
\end{equation}
with
\begin{equation}
V(a)=\frac{A_1(a)+A_2(a)}{2}-(2\pi a\sigma(a))^2-\left(\frac{A_1(a)-A_2(a)}{8\pi a\sigma(a)}\right)^2 .
\label{eq:potential}
\end{equation}
The function $V(a)$ can be understood as an effective potential. A static solution satisfies $V(a_0)=V'(a_0)=0$ and is linearly stable under radial perturbations if
$V''(a_0)>0$. Following \cite{BEFA2025}, we adopt a linearized equation of state at the shell,
\begin{equation}
p-p_0=\eta(\sigma-\sigma_0)+\mathcal{O}\!\left((\sigma-\sigma_0)^2\right),
\label{eq:eos}
\end{equation}
where $\eta$ is a constant that can be interpreted as the squared sound speed when $0\le \eta<1$. This closes the system and allows one to map the stability regions in the relevant  parameter space. We can see from Eq. (\ref{conserv}) that $a\sigma '(a) = -2(\sigma(a)+p(a))$ and, after some algebra, the second derivative of the potential finally takes the form
\begin{align}
V''(a_0)&= - \frac{\sqrt{A_2(a_0)} A_1''(a_0)-\sqrt{A_1(a_0)} A_2''(a_0)}{ \sqrt{A_1(a_0)}-\sqrt{A_2(a_0)}}
+\frac{A_2(a_0)^{3/2} A_1'(a_0)^2-A_1(a_0)^{3/2} A_2'(a_0)^2}{2 A_1(a_0) A_2(a_0) \left(\sqrt{A_1(a_0)}-\sqrt{A_2(a_0)}\right)} \nonumber \\ 
& +(2 \eta +1)\left( -\frac{ \sqrt{A_2(a_0)}\left( a_0 A_1'(a_0)- 2 A_1(a_0)\right) }{ a_0^2  \left( \sqrt{A_1(a_0)}-\sqrt{A_2(a_0)}\right)}
+\frac{\sqrt{A_1(a_0)}\left( a_0 A_2'(a_0)-2 A_2(a_0)\right) }{ a_0^2  \left( \sqrt{A_1(a_0)}-\sqrt{A_2(a_0)}\right)}\right) .
\label{potD2VSinSigmaP}
\end{align}
We can find the stable configurations under radial perturbations from the analysis of the sign of $V''(a_0)$.

\section{Charged vacuum bubbles: representative stability regions}

We consider ``charged bubbles'' with a Minkowski interior,
\begin{equation}
A_1(r)=1,
\end{equation}
and an exterior given by one of the UG charged solutions presented above (Case A or Case B), that is\ $A_2(r)$ given by \eqref{eq:caseA_A} or \eqref{eq:caseB_A}.
The shell radius $a_0$ is chosen larger than the event horizon radius $r_h$ of the original exterior solution and, in de Sitter-like cases, smaller than the cosmological horizon radius $r_c$. This yields a regular vacuum core and a global spacetime without event horizons or central singularities; a cosmological horizon may still exist outside the shell when $\Lambda_0>0$.

\begin{figure*}[t!]
\centerline{\includegraphics[width=0.95\textwidth]{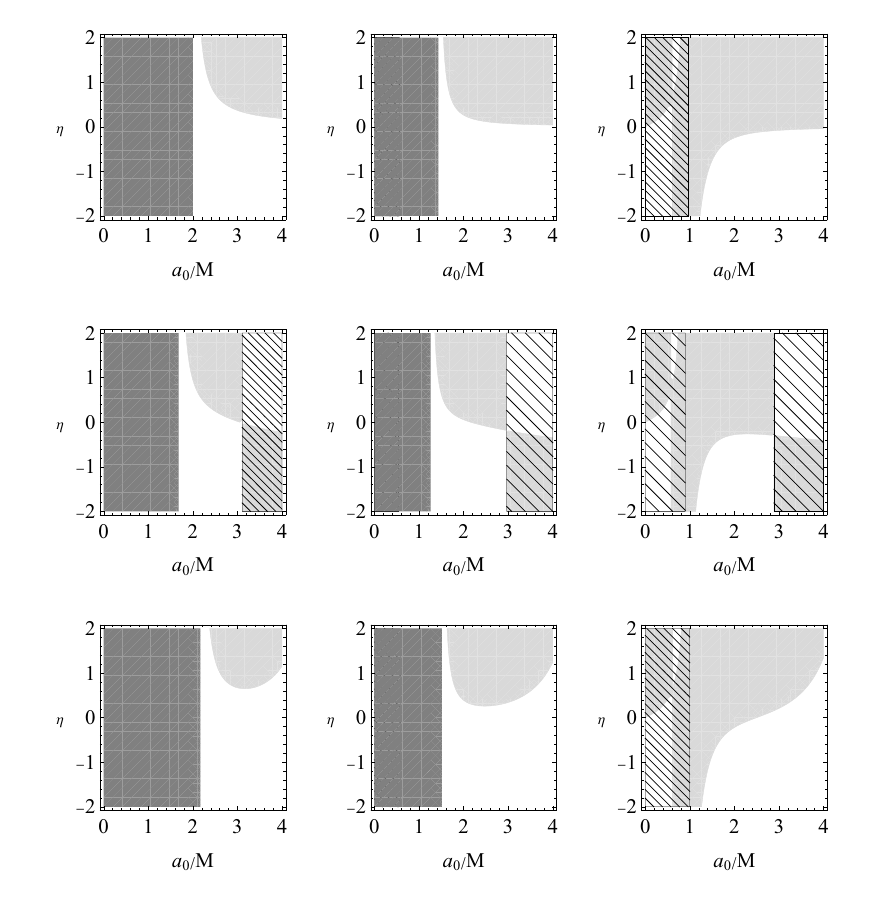}}
\caption{Stability regions for charged vacuum bubbles in the $(a_0/M,\eta)$ plane in the GR scenario, which is recovered when $\Lambda_1 =0$ in the UG exterior solutions. In all plots, the light gray regions denote stable configurations, the dashed ones violate WEC, and the dark gray zones are non-physical. From top to bottom, the rows display $\Lambda_0 M^2=0$, $\Lambda_0M^2=-0.2$, and $\Lambda_0M^2=0.05$, while the critical values of charge are $Q_c/M=1$, $Q_c/M=0.97$, and $Q_c/M=1.01$. From left to right, the columns show $Q=0$, $|Q|=0.9Q_c$, and $|Q|=1.1Q_c$. In the bottom row, there is a cosmological horizon at $r_c=6.43$, $r_c=6.54$, and $r_c=6.60$ (not shown). }
\label{fig1}
\end{figure*}

\begin{figure*}[t!]
\centerline{\includegraphics[width=0.95\textwidth]{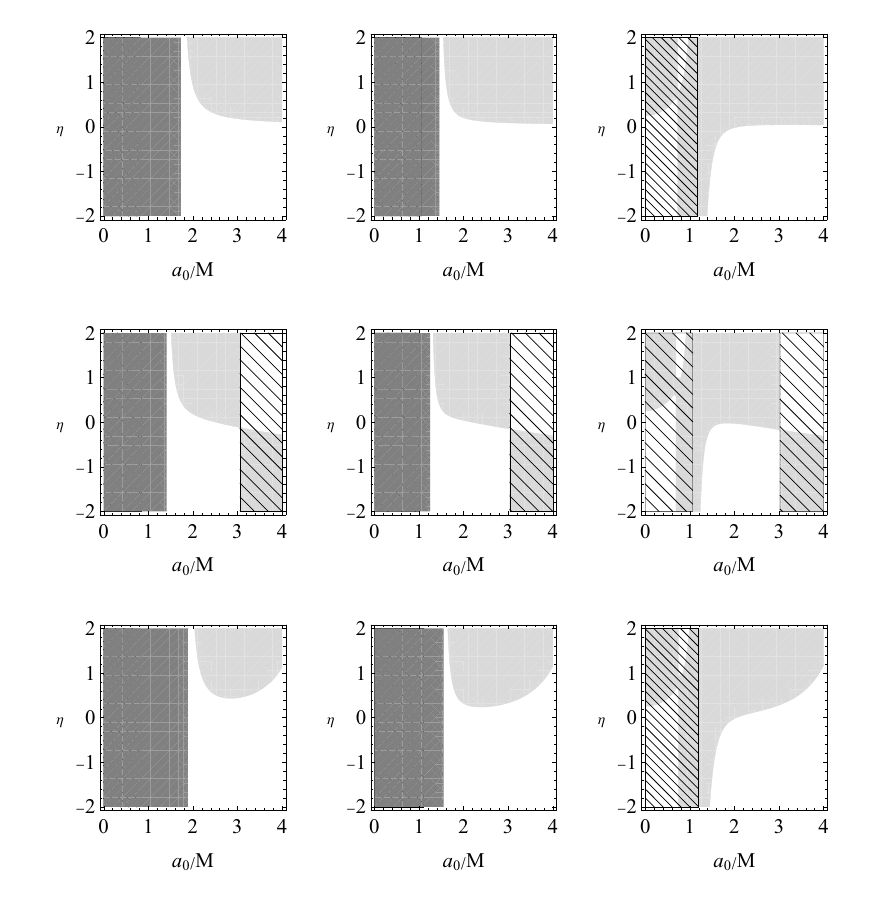}}
\caption{Analogous to Fig. \ref{fig1} for UG exterior Case A, with  $p=-5$ and $\Lambda_1 M^{p+2}=-0.4$. From top to bottom, the rows display $\Lambda_0 M^2=0$, $\Lambda_0M^2=-0.2$, and $\Lambda_0M^2=0.05$, while the critical values of charge are $Q_c/M=0.55$, $Q_c/M=0.41$, and $Q_c/M=0.59$. From left to right, the columns show $Q=0$, $|Q|=0.9Q_c$, and $|Q|=1.1Q_c$. In the bottom row, the cosmological horizon is located at $r_c=6.45$, $r_c=6.49$, and $r_c=6.50$ (not shown). }
\label{fig2}
\end{figure*}

\begin{figure*}[t!]
\centerline{\includegraphics[width=0.95\textwidth]{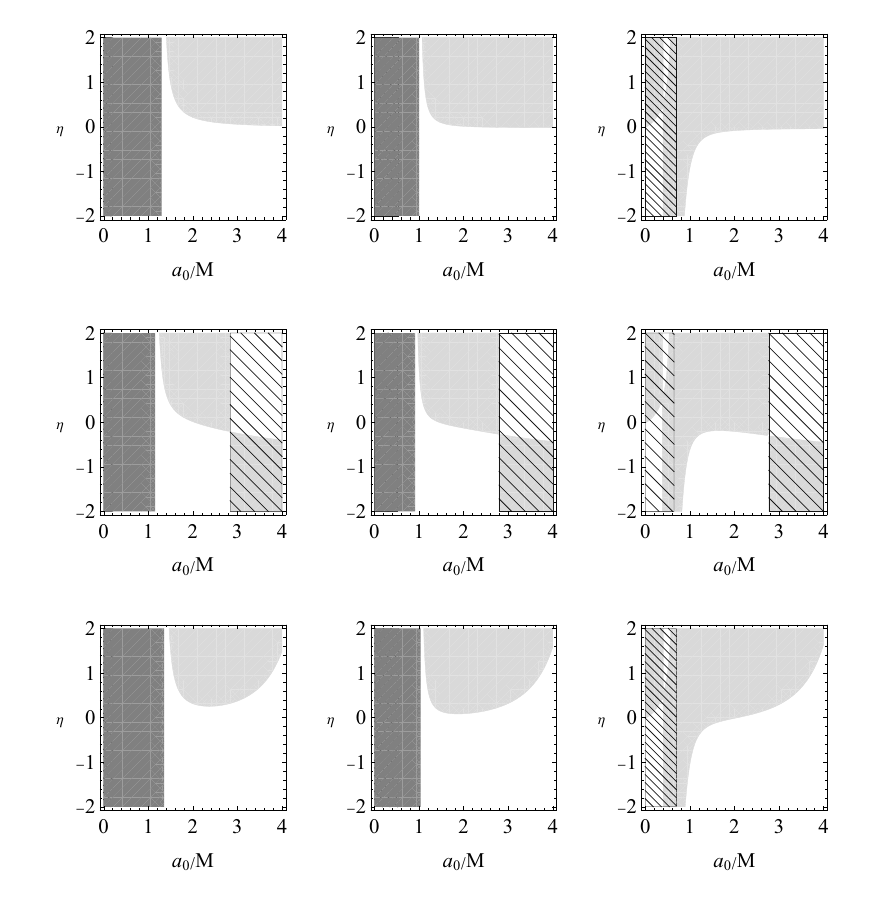}}
\caption{Analogous to Fig. \ref{fig1} for the UG exterior Case A, with  $p=-7/2$ and $\Lambda_1 M^{p+2}=0.05$. From top to bottom, the rows display $\Lambda_0 M^2=0$, $\Lambda_0M^2=-0.2$, and $\Lambda_0M^2=0.05$, while the critical values of charge are $Q_c/M=0.495$, $Q_c/M= 0.474$, and $Q_c/M=0.501$. From left to right, the columns show $Q=0$, $|Q|=0.9Q_c$, and $|Q|=1.1Q_c$. In the bottom row, the cosmological horizon is located at $r_c=6.70$, $r_c=6.72$, and $r_c=6.73$ (not shown).}
\label{fig3}
\end{figure*}

\begin{figure*}[t!]
\centerline{\includegraphics[width=0.95\textwidth]{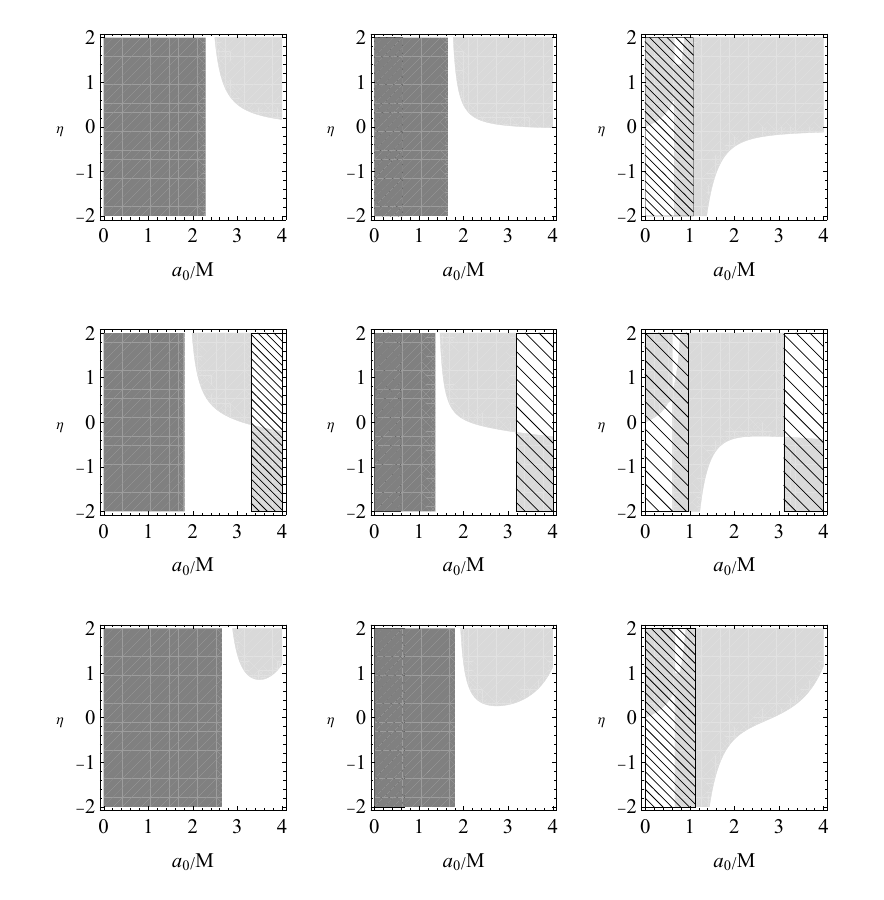}}
\caption{Analogous to Fig. \ref{fig1} for the UG exterior Case B, with $\Lambda_1 M^{-2}=0.4$ and $b/M=1$. From top to bottom, the rows display $\Lambda_0 M^2=0$, $\Lambda_0M^2=-0.2$, and $\Lambda_0M^2=0.05$, while the critical values of charge are $Q_c/M=1.04$, $Q_c/M=1.00$, and $Q_c/M=1.06$. From left to right, the columns show $Q=0$, $|Q|=0.9Q_c$, and $|Q|=1.1Q_c$. In the bottom row, the cosmological horizon is located at $r_c=5.68$, $r_c=5.89$, and  $r_c=5.98$ (not shown).}
\label{fig4}
\end{figure*}

We classify solutions by their stability, determined from the sign of $V''(a_0)$, and by their matter content at the shell, through the weak energy condition (WEC): $\sigma_0\ge 0$ and $\sigma_0+p_0\ge 0$. A key role is played by the critical charge $Q_c$ at which the horizon structure of the exterior geometry changes \cite{BEFA2025}. We present our results graphically for some representative scenarios in Figs. \ref{fig1}--\ref{fig4}. These figures show two examples with an exterior UG solution Case A with different values of the power $p$ and another one corresponding to UG Case B; the GR counterpart is also presented for comparison. In all plots, the light gray regions denote stable configurations in the plane  $(a_0/M,\eta)$; the dashed ones violate WEC and the dark gray regions are non-physical because $a_0$ is not larger than the event horizon radius.

Figure \ref{fig1} shows the stability zones for the GR case (which is equivalent to taking $\Lambda_1=0$ in the UG solutions). The rows display, from top to bottom, the results for $\Lambda_0 M^2=0$, $\Lambda_0 M^2=-0.2$, and $\Lambda_0 M^2=0.05$; the corresponding critical values of the charge are $Q_c/M=1$, $Q_c/M=0.97$, and $Q_c/M=1.01$, respectively. Columns are organized by charge, from left to right, $Q=0$, $|Q|=0.9Q_c$, and $|Q|=1.1Q_c$. In the bottom row (outside the displayed range), there is a cosmological horizon located, from left to right, at $r_c=6.43$, $r_c=6.54$, and $r_c=6.60$, respectively.

Figures \ref{fig2} and \ref{fig3} show the results for the UG Case A, for two distinct values of $p$. For Fig. \ref{fig2}, with $p=-5$, a negative value of $\Lambda_1$ is needed; 
in particular, we take $\Lambda_1 M^{p+2}=-0.4$. The rows display, from top to bottom, the results for $\Lambda_0 M^2=0$, $\Lambda_0 M^2=-0.2$, and $\Lambda_0 M^2=0.05$; the corresponding critical values of the charge are $Q_c/M=0.55$, $Q_c/M=0.41$, and $Q_c/M=0.59$, respectively. The cosmological horizon in the last row (outside the displayed range) is located, from left to right, at $r_c=6.45$, $r_c=6.49$, and $r_c=6.50$. For Fig. \ref{fig3}, with $p=-7/2$, a positive value of  $\Lambda_1$ is required, then we adopt $\Lambda_1 M^{p+2}= 0.05$. The rows display, from top to bottom, the results for $\Lambda_0 M^2=0$, $\Lambda_0 M^2=-0.2$, and $\Lambda_0 M^2=0.05$; the corresponding critical values of the charge are $Q_c/M=0.495$, $Q_c/M=0.474$, and $Q_c/M=0.501$, respectively. The cosmological horizon in the last row is located, from left to right, at $r_c=6.70$, $r_c=6.72$, and $r_c=6.73$. In case $p=-7/2$ the UG contribution, in both the metric and the electromagnetic field, decays more slowly with the radial coordinate than for $p=-5$.

In Fig. \ref{fig4}, Case B with $\Lambda_1/M^2=0.4$ and $b/M=1$ is shown. The rows display, from top to bottom, the results for $\Lambda_0 M^2=0$, $\Lambda_0 M^2=-0.2$, and $\Lambda_0 M^2=0.05$; the corresponding critical values of charge are $Q_c/M=1.04$, $Q_c/M=1.00$, and $Q_c/M=1.06$. There is also a cosmological horizon in the last row  (outside the displayed range) located, from left to right, at $r_c=5.68$, $r_c=5.89$, and $r_c=5.98$. 

The comparison of UG with GR shows: (i) the critical charge $Q_c/M$ is smaller in UG Case A and larger in UG Case B than in GR; (ii) for fixed $(\Lambda_0M^2,|Q|/M)$, stability regions are typically slightly larger in the UG examples displayed; and (iii) departures from GR grow with $|\Lambda_1|$. These trends arise because $\Lambda_1$ modifies both the electric field and the metric function in the non-conservative UG solutions, changing $\sigma_0$, $p_0$, and ultimately $V''(a_0)$.

Let us analyze the WEC at the thin shell. For the Minkowski interior considered here, the WEC can be expressed directly in terms of the exterior metric function. From Eqs.  \eqref{eq:sigma0} and \eqref{eq:p0}, we obtain
\begin{equation*}
\sigma_0=\frac{1-\sqrt{A_2(a_0)}}{4\pi a_0},
\qquad
\sigma_0+p_0=
\frac{1-\sqrt{A_2(a_0)}}{8\pi a_0}
+\frac{A_2'(a_0)}{16\pi\sqrt{A_2(a_0)}}.
\end{equation*}
Hence, the WEC requires to satisfy
\begin{equation}
A_2(a_0)\leq 1
\qquad
\mathrm{and}
\qquad
a_0A_2'(a_0)+2\sqrt{A_2(a_0)}\left( 1-\sqrt{A_2(a_0)}\right) \geq0.
\end{equation}
Note that these inequalities are independent of $\eta$, which only enters the linearized stability analysis. They also explain the main qualitative features displayed in Figs. \ref{fig1}--\ref{fig4}, in which violating WEC configurations correspond to the dashed regions. For the asymptotically AdS configurations considered here, the WEC necessarily fails at sufficiently large shell radii, whereas in the de Sitter case it is violated sufficiently close to the cosmological horizon. Although the charge does not appear explicitly in these expressions, its dependence is encoded in $A_2$ and $A_2'$. In particular, crossing the critical charge $Q_c$ changes the horizon structure of the exterior geometry and hence the allowed range of $a_0$, which explains the qualitatively different WEC domains displayed for subcritical and supercritical charges. In all figures, we find several features shared by GR and all the UG examples considered here. When $\Lambda_0 = 0$, if $|Q|<Q_c$ the WEC is always fulfilled for any possible radius $a_0/M$, while for $|Q|>Q_c$ it is not satisfied for small values of $a_0/M$. When $\Lambda_0 < 0$, if $|Q|<Q_c$ the WEC is not fulfilled for values larger than $a_0/M \sim 3$, while for $|Q|>Q_c$ it is not satisfied for small values of $a_0/M$ and for values larger than $a_0/M \sim 3$. When $\Lambda_0 > 0$, if $|Q|<Q_c$ the WEC is not fulfilled in a region close to the cosmological horizon (not shown), while for $|Q|>Q_c$, an additional zone where WEC is not satisfied for small values of $a_0/M$ appears.  In all the scenarios considered, there exist regions in the $(a_0/M,\eta )$ plane where the configurations are stable with the matter at the shell also satisfying WEC.

\section{Conclusions}

We have summarized the construction of spherical thin shells within UG by using the corresponding junction conditions and we have outlined the stability analysis under radial perturbations for a large family of spacetimes. As an application, we have considered charged vacuum bubbles with Minkowski interiors and non-conservative UG charged exteriors (Cases A and B). The resulting spacetimes avoid the presence of event horizons and central singularities, while admitting stable configurations for suitable parameter values. For Case A, we have adopted two values of the power $p$. The choice $p=-7/2$ provides a qualitatively different radial behavior from the $p=-5$ case, since the UG contribution to both the metric function and the electric field decays more slowly than the corresponding charge term. In comparison with GR, we find shifts in the critical charge controlling the horizon structure and modest enlargements of stability regions for representative choices of parameters. We have presented a detailed analysis of the matter at the thin shell by studying the WEC, for the different exterior solutions. All the configurations show a region in the parameter space where stability and WEC are simultaneously satisfied.

These results, which extend the previous studies in the literature, illustrate how the non-conservative bulk allowed by UG can modify the matter content and the stability properties of the thin shells.

\section*{Acknowledgements}

This work was supported by CONICET. We thank J.\ C.\ Fabris for useful discussions on charged UG solutions.

\end{document}